\documentclass[twocolumn]{aastex61}
\usepackage{CJK} 

\usepackage{amsmath}
\usepackage{amssymb}
\usepackage{epsfig}
\usepackage{apjfonts}
\usepackage{natbib}
\usepackage{hyperref}
\usepackage{epstopdf}
\usepackage{verbatim}
\usepackage{enumitem}
\usepackage{color}
\setlist[enumerate]{itemsep=-1mm}
\usepackage{soul, xcolor}
\setstcolor{blue}

\usepackage{mathtools}

\usepackage{bigints}

\usepackage{tikz}

\makeatletter
\global\let\tikz@ensure@dollar@catcode=\relax
\makeatother

\newcommand{\Teffstar}{\mbox{$T_{\rm eff,\star}$}}
\newcommand{\Rstar}{\mbox{$R_{\star}$}}
\newcommand{\Mstar}{\mbox{$M_{\star}$}}

\newcommand{\logLbolstar}{\mbox{$\log(L_{\rm bol,\star}/\Lsun)$}}
\newcommand{\Teff}{\mbox{$T_{\rm eff}$}}
\newcommand{\R}{\mbox{$R$}}
\newcommand{\M}{\mbox{$M$}}
\newcommand{\Lbol}{\mbox{$L_{\rm bol}$}}

\newcommand{\kms}{\hbox{km\,s$^{-1}$}}
\newcommand{\vsini}{\mbox{$v\sin{i}$}}
\newcommand{\veq}{\mbox{$v_{\rm eq}$}}

\newcommand{\Lsun}{\mbox{$L_{\odot}$}}
\newcommand{\Msun}{\hbox{$M_{\odot}$}}
\newcommand{\Rsun}{\hbox{$R_{\odot}$}}
\newcommand{\Mjup}{\mbox{$M_{\rm Jup}$}}
\newcommand{\Rjup}{\mbox{$R_{\rm Jup}$}}

\newcommand{\logg}{\mbox{$\log(g)$}}
\def\sun{\hbox{$\odot$}}
\def\earth{\hbox{$\oplus$}}

\newcommand{\Sinit}{\mbox{$S_{\rm init}$}}
\newcommand{\kbb}{\mbox{$k_{\rm B}$\,baryon$^{-1}$}}

\newcommand{\etal}{et al.}

\newcommand{\eg}{e.g.}

\newcommand{\GALEX}{{\sl GALEX}}

\newcommand{\logLbol}{\mbox{$\log(L_{\rm bol}/\Lsun)$}}

\received{May 8, 2021; Revised July 1, 2021; Accepted July 4, 2021}
\submitjournal{ApJ Letters}

\begin{document}


\begin{CJK*}{UTF8}{gbsn}

\title{The Second Discovery from the COol Companions ON Ultrawide orbiTS (COCONUTS) Program: \\ A Cold Wide-Orbit Exoplanet around a Young Field M Dwarf at 10.9 pc}

\author[0000-0002-3726-4881]{Zhoujian Zhang (张周健)}
\affiliation{Institute for Astronomy, University of Hawaii at Manoa, Honolulu, HI 96822, USA}

\author[0000-0003-2232-7664]{Michael C. Liu}
\affiliation{Institute for Astronomy, University of Hawaii at Manoa, Honolulu, HI 96822, USA}

\author[0000-0002-9879-3904]{Zachary R. Claytor}
\affiliation{Institute for Astronomy, University of Hawaii at Manoa, Honolulu, HI 96822, USA}

\author[0000-0003-0562-1511]{William M. J. Best}
\affiliation{The University of Texas at Austin, Department of Astronomy, 2515 Speedway, C1400, Austin, TX 78712, USA}

\author[0000-0001-9823-1445]{Trent J. Dupuy}
\affiliation{Institute for Astronomy, University of Edinburgh, Royal Observatory, Blackford Hill, Edinburgh, EH9 3HJ, UK}

\author[0000-0001-5016-3359]{Robert J. Siverd}
\affiliation{Gemini Observatory/NSF's NOIRLab, 670 N. A`ohoku Place, Hilo, HI, 96720, USA}

\begin{abstract}
We present the identification of the COCONUTS-2 system, composed of the M3 dwarf L~34-26 and the T9 dwarf WISEPA~J075108.79$-$763449.6. Given their common proper motions and parallaxes, these two field objects constitute a physically bound pair with a projected separation of $594''$ (6471\,au). The primary star COCONUTS-2A has strong stellar activity (H$\alpha$, X-ray, and UV emission) and is rapidly rotating ($P_{\rm rot} = 2.83$~days), from which we estimate an age of $150-800$~Myr. Comparing equatorial rotational velocity derived from the TESS light curve to spectroscopic \vsini, we find COCONUTS-2A has a nearly edge-on inclination. The wide exoplanet COCONUTS-2b has an effective temperature of \Teff$=434 \pm 9$~K, a surface gravity of $\log{g} = 4.11^{+0.11}_{-0.18}$~dex, and a mass of \M$=6.3^{+1.5}_{-1.9}$~\Mjup\ based on hot-start evolutionary models, leading to a mass ratio of $0.016^{+0.004}_{-0.005}$ for the COCONUTS-2 system. COCONUTS-2b is the second coldest (after WD~0806$-$661B) and the second widest (after TYC 9486-927-1~b) exoplanet imaged to date. Comparison of COCONUTS-2b's infrared photometry with ultracool model atmospheres suggests the presence of both condensate clouds and non-equilibrium chemistry in its photosphere. Similar to 51~Eri~b, COCONUTS-2b has a sufficiently low luminosity ($\log{(L_{\rm bol}/L_{\odot})} = -6.384 \pm 0.028$~dex) to be consistent with the cold-start process that may form gas-giant (exo)planets, though its large separation means such formation would not have occurred in situ. Finally, at a distance of 10.9~pc, COCONUTS-2b is the nearest imaged exoplanet to Earth known to date.
\end{abstract}

\section{Introduction}
\label{sec:introduction}

Direct imaging of exoplanets provides a valuable window into the atmospheres, formation, and evolution of gas-giant planets \citep[\eg,][]{2016PASP..128j2001B}.   In addition to photometry and spectroscopy of the exoplanets themselves, characterization of the host stars is also critical in such investigations, since system properties such as birth environment, stellar insolation, and host mass, age, and metallicity are central to understanding the past, present, and future of the exoplanets.  Exoplanetary systems with wide-separation planets are particularly useful, given the relative ease with which both the host star and the exoplanets can be characterized, in comparison to smaller-separation exoplanets where the host's starlight can impede direct observations.
As part of our ongoing COol Companions ON Ultrawide orbiTS (COCONUTS) program to find wide separations planetary-mass and substellar companions to stars \citep[e.g.,][]{2020ApJ...891..171Z}, we have established the physical association of the active M~dwarf L~34-26 (PM~J07492$-$7642, TYC~9381-1809-1; hereinafter COCONUTS-2A) with the T9 dwarf WISEPA~J075108.79$-$763449.6 (hereinafter, COCONUTS-2b). This Letter presents characterization of this system, with astrometric, spectrophotometric, and physical properties summarized in Table~\ref{tab:info}.

\begin{figure*}[t]
\begin{center}
\includegraphics[height=6in]{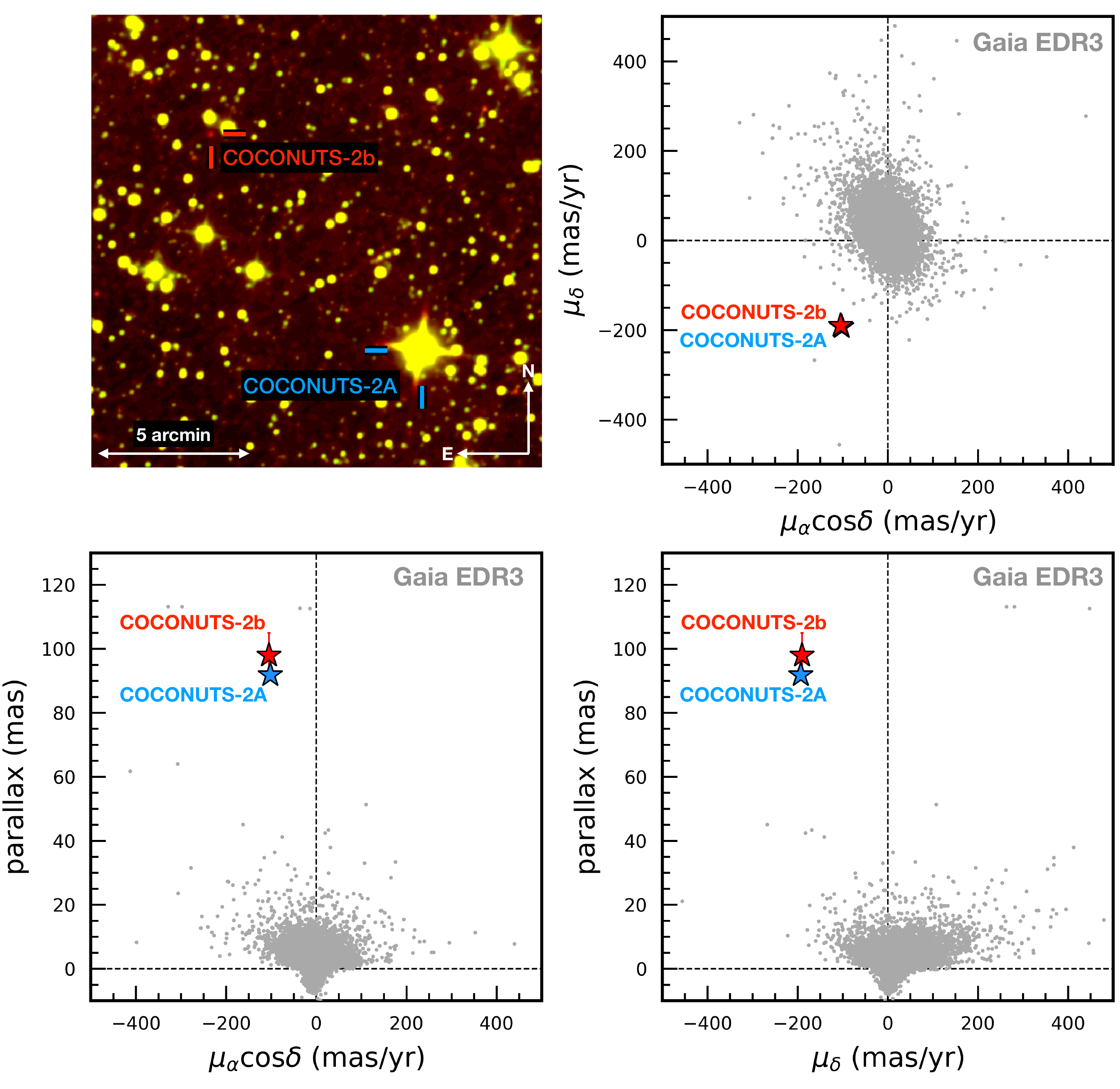}
\caption{{\it Upper Left}: The COCONUTS-2 system in the bi-color AllWISE image ($W1$: green, $W2$: red). COCONUTS-2b stands out from the field with a very red $W1-W2$ color. The A and b components are separated by 594\arcsec, corresponding to 6471~au at the primary's distance. {\it Upper Right \& Bottom}: Proper motion and parallax of COCONUTS-2A (blue; Gaia EDR3) and COCONUTS-2b \citep[red;][]{2019ApJS..240...19K}. Overlaid are the Gaia EDR3 sources within a radius of $5^{\circ}$ (grey), with typical proper motion and parallax uncertainties of $0.3$~mas~yr$^{-1}$ and $0.2$~mas, respectively. The physical association between the A and b components is clearly indicated by their common proper motions and parallaxes.}
\label{fig:pair}
\end{center}
\end{figure*}

\section{The COCONUTS-2 System}
\label{sec:system}

As part of the COCONUTS program, we examined astrometry of 440 free-floating T5$-$Y1 dwarfs in the UltracoolSheet\footnote{\url{http://bit.ly/UltracoolSheet}} to search for their co-moving primary stars from Gaia~EDR3 \citep[][]{2016AandA...595A...1G, 2020arXiv201201533G}. We identify co-moving pairs if: (1) the projected separation between the primary star and the companion is $<10^{4}$~au; (2) the difference between the two components' parallaxes are within $30\%$ of the primary star's parallax; and (3) the vector difference between the two components' proper motions are within $30\%$ of the primary star's total proper motion \citep[e.g.,][]{2012ApJS..201...19D}. The COCONUTS-2 system was identified in this process. 

COCONUTS-2A has a parallax $= 91.826 \pm 0.019$~mas \citep[$= 10.888 \pm 0.002$~pc;][]{2021AJ....161..147B} and a proper motion of $(\mu_{\alpha}\cos{\delta}, \mu_{\delta}) = (-102.15\pm0.02,\ -192.92 \pm 0.02)$~mas~yr$^{-1}$ from Gaia~EDR3, measured from 25 visibility periods with a Renormalised Unit Weight Error of 1.07. COCONUTS-2b has a parallax $= 97.9 \pm 6.7$~mas ($= 10.2 \pm 0.7$~pc) and a proper motion of $(\mu_{\alpha}\cos{\delta}, \mu_{\delta}) = (-104.8\pm2.8,\ -189.7 \pm 4.5)$~mas~yr$^{-1}$ as measured by \cite{2019ApJS..240...19K} using 18 epochs of ground-based astrometry spanning 5.3~years. The physical association between the A and b components is clearly indicated by their common distances and proper motions (Figure~\ref{fig:pair}), with a consistency of $0.98\sigma$ and $1.07\sigma$, respectively. The angular separation between these two components is $594''$, corresponding to a projected separation of $6471$~au at the primary star's distance.

To estimate the chance alignment probability of the COCONUTS-2 system, we simulate the kinematics of field ultracool dwarfs and identify synthetic objects that would be identified as COCONUTS-2A's companions by our aforementioned criteria. We generate $10^{7}$ ensembles of ultracool dwarfs for this test. In each ensemble, we assume objects are uniformly distributed within a 20~pc-radius sphere centered on Earth with a space density of $1.33 \times 10^{-2}$~pc$^{-3}$ \citep[measured for T5$-$Y1 dwarfs by][]{2021ApJS..253....7K}, leading to random Galactic $XYZ$ positions for a total of 446 synthetic objects. We generate random $UVW$ space motions of these objects by assuming 3D normal distributions with the mean and standard deviation of $U = 2.7 \pm 31.6$~\kms, $V = -5.1 \pm 27.0$~\kms, $W = 1.1 \pm 16.5$~\kms\ as determined by \cite{2015ApJS..220...18B} for M and L dwarfs within 20~pc. We then convert these synthetic $XYZUVW$ into equatorial coordinates, proper motions, parallaxes, and radial velocities, finding that 24 synthetic objects would be identified as a companion of COCONUTS-2A among all $10^{7}$ ensembles. Given that our full search examined 440 T5$-$Y1 dwarfs, the probability that our search would result in a single chance alignment ``discovery'' like COCONUTS-2b is thus $24 \times 10^{-7} \times 440 = 10^{-3}$.

\section{COCONUTS-2A: The M dwarf Primary Star}
\label{sec:primary}

\subsection{Physical Properties \label{subsec:phys_primary}}
COCONUTS-2A is a strong H$\alpha$, X-ray, and UV-emitting M3 dwarf. \cite{2019AandA...629A..80H} measured [Fe/H]$= 0.00 \pm 0.08$~dex from VLT/UVES spectra, which we find is consistent with the metallicity of $-0.05 \pm 0.17$~dex we derive from the $V-K$ color using the \cite{2012A&A...538A..25N} calibration. \cite{2014MNRAS.443.2561G} derived an effective temperature (\Teffstar) of $3406 \pm 69$~K, by comparing this star's optical spectrum with PHOENIX BT-Settl models. Using the empirical relations of \cite{2015ApJ...804...64M} and \cite{2019ApJ...871...63M}, we  compute $\Rstar=0.388\pm0.011$\,\Rsun, $\Mstar=0.37 \pm 0.02$\,\Msun, and \logLbolstar$=-1.732\pm0.011$~dex for COCONUTS-2A, with calibration errors and measurement uncertainties propagated in an analytic fashion. 

\cite{2019AandA...629A..80H} reported the system is a possible spectroscopic binary based on initial examination of their UVES spectra, though they did not corroborate this in their quantitative analysis.  High-resolution optical spectroscopy has also been obtained by \cite{2006AandA...460..695T} and \cite{2019AJ....157..234S}, neither of whom reported evidence for spectroscopic binarity and measured consistent radial velocities (0.9$\pm$0.3~\kms\ from ESO 1.5m/FEROS and 1.2$\pm$0.6~\kms\ from Magellan/MIKE, respectively).  Thus we conclude that COCONUTS-2A is unlikely to be a spectroscopic binary.

Combining our estimated \Rstar\ with the rotation period derived from the TESS light curve (see below), we compute an equatorial rotational velocity of \veq$=6.9\pm0.7$~\kms\ for COCONUTS-2A. This velocity is very close to the spectroscopic \vsini$=8.1\pm1.2$~\kms\ by \cite{2006AandA...460..695T} and $7.4\pm0.4$~\kms\ by \cite{2019AandA...629A..80H}, leading to $\sin{i} = 1.17^{+0.23}_{-0.20}$ and $1.07^{+0.14}_{-0.12}$, respectively, by assuming \veq\ and \vsini\ follow normal distributions with a truncation at 0. These suggest COCONUTS-2A has a nearly $90^{\circ}$ (edge-on) inclination. The star does not host any known or candidate transiting planets based on ExoFOP.\footnote{\url{https://exofop.ipac.caltech.edu/index.php}}

\subsection{Age \label{subsec:age}}
We estimate the star's age using spectroscopic, activity, rotation, kinematic, and photometric properties.

{\it Lithium.}  Li~$\lambda6708$ absorption is absent in high-resolution spectra ($R\sim35000-50000$) with an upper limit on the equivalent width (EW) $<$50~m\AA\  \citep{2019AJ....157..234S}. Assuming $99\%$ of the object's initial Li abundance is depleted, the \cite{1997A&A...327.1039C} models predict COCONUTS-2A is likely $>30$~Myr, given its M3 spectral type and $K_{S}$-band absolute magnitude.

{\it Alkali Lines.} The Na~I ($\lambda8183/8195$) and K~I ($\lambda7665/7699$) doublets are sensitive to  surface gravity and thus a possible youth indicator for M dwarfs. \cite{2014AJ....147...85R} measured an Na~I index $= 1.18$ and EW$_{\rm K7699} = 1.0$~\r{A} for COCONUTS-2A, which suggest an age of $\gtrsim100$~Myr given the star's $V-K_{S}=4.70\pm0.07$~mag, though Reidel \etal\ advise caution in the interpretation of these lines at $V-K_{S}<5$. 

{\it H$\alpha$ Emission.} COCONUTS-2A exhibits strong H$\alpha$ emission, with EW$_{\rm H\alpha} = 2.3 - 8.0$~\r{A} from the literature (Table~\ref{tab:info}). Adopting EW$_{\rm H\alpha} = 4.1 \pm 1.2$~\AA\ based on the median and 16th-to-84th percentiles, we compute $\log_{10}{(L_{\rm H\alpha}/L_{\rm bol})} = \log_{10}{({\rm EW_{H\alpha} \times \chi})} = -3.74 \pm 0.13$~dex, with $\chi=(4.4872\pm0.4967)\times10^{-5}$ from \cite{2014ApJ...795..161D}. To estimate the stellar age, we adapt the \cite{2021AJ....161..277K} H$\alpha$ activity-age relation into a Bayesian framework. We use an MCMC process to derive the stellar age ($t$) posterior using a log-likelihood function:
\begin{eqnarray}
\ln{\mathcal{L}(t)} &=& -\frac{\left[\log(L_{\rm H\alpha}/L_{\rm bol})_{\rm model} - \log(L_{\rm H\alpha}/L_{\rm bol})_{\rm obs}\right]^{2}}{2 \times (\sigma_{\rm obs}^2 + \sigma_{V}^{2})}\ ,
\end{eqnarray}
where $\log(L_{\rm H\alpha}/\Lbol)_{\rm obs} = -3.74$~dex and $\sigma_{\rm obs} = 0.13$~dex. 
We use the best-fit model parameters from \cite{2021AJ....161..277K} to compute $\log(L_{\rm H\alpha}/L_{\rm bol})_{\rm model}$ as a function of $t$, with $\sigma_{V} = 0.22$~dex. Assuming a uniform prior of [1.5~Myr, 10~Gyr] in $t$, we run {\it emcee} \citep{2013PASP..125..306F} with 20 walkers and 5000 iterations (with the first 100 ``burn-in'' iterations excluded) and obtain $t = 750^{+800}_{-500}$~Myr. This age is consistent with the dynamical model-based activity lifetime of $<2 \pm 0.5$~Gyr for M3 dwarfs \citep{2008AJ....135..785W}.

{\it X-ray and UV Emission.} \cite{2019AJ....157..234S} measured an X-ray luminosity of $L_{X} = 28.92 \pm 0.02$~dex using the second ROSAT all-sky survey. This $L_{X}$ is similar to those of ONC ($\approx1$~Myr), Pleiades ($\approx112$~Myr), and Hyades ($\approx750$~Myr) members \citep[e.g.,][]{2005ApJS..160..390P} but much larger than field-age dwarfs. We derive $\log{R_{X}} = \log{(L_{X}/L_{\rm bol})} = -2.93 \pm 0.02$~dex, which falls in the saturation regime of the X-ray activity-age relation. We also compute the ratio between X-ray and 2MASS $J$-band fluxes, $\log{(f_{X}/f_{J})}=\log{(f_{X})}+0.4J+6.3=-1.97\pm0.02$~dex, which is consistent with those of Pleiades members and higher than most Hyades members \citep[e.g.,][]{2009ApJ...699..649S}. \cite{2019AJ....157..234S} also converted COCONUTS-2A's \GALEX\ UV photometry to flux ratios of $\log{(f_{\rm NUV}/f_{J})}=-3.94\pm0.02$~dex and $\log{(f_{\rm FUV}/f_{J})}=-4.64\pm0.06$~dex, which we find are higher than field dwarfs with comparable stellar masses but similar to members of the Hyades and younger associations \cite[see Figures~3--4 of][]{2018AJ....155..122S}. The X-ray and UV activity thus suggests an age of $\lesssim 750$~Myr.

{\it Rotation.} We used a Lomb-Scargle periodogram to estimate the star's rotation period, based on TESS data from sectors 3, 4, 6, 7, 10, 11, 12, 13, 27, 30, 33, 34, and 36, spanning a range of 920 days. We queried the two-minute cadence, SPOC PDCSAP light curves from MAST using \texttt{lightkurve} \citep{2018ascl.soft12013L}. We found a rotation period of $P_{\rm rot}=$2.8299$\pm$0.0025 (rand) $\pm$0.2830 (sys) days, where the random uncertainty was estimated by fitting a Gaussian to the main periodogram peak using least-squares minimization. We assumed a 10\% systematic uncertainty to account for the possibility of surface differential rotation. The peak false alarm probability was vanishingly small, and the only other significant peak was the half-period alias, which we ruled out upon visual inspection of the light curve. The light curve amplitude was 2.2\%$\pm$0.3\%, which we obtained by finding the difference between the 95th and 5th percentiles of flux for each sector and computing the median over all sectors. Our TESS-based rotation period is also consistent with the periods of 2.827 and 2.829 days measured using the ASAS and ASAS-SN photometry by \cite{2012AcA....62...67K} and \cite{2019MNRAS.486.1907J}, respectively. The primary star's $P_{\rm rot}$ lines up with members of Pleiades and younger associations ($\lesssim 60$~Myr) and lies at the short-period end of Praesepe members \citep[$\sim 800$~Myr; see][]{2018AJ....155..196R}. We therefore draw a rotation-based age estimate of $\lesssim 800$~Myr.

{\it Kinematics.} Based on Gaia~EDR3 astrometry and the radial velocity measured by \cite{2019AJ....157..234S}, we find COCONUTS-2A is not associated with any young moving group using BANYAN~$\Sigma$ \citep[][]{2018ApJ...856...23G}.  The low tangential velocity ($11.276 \pm 0.002$~\kms) of the star is consistent with being a young field object. However, the object's space motion $UVW = (+6.4 \pm 0.2, +4.5 \pm 0.5, -8.2 \pm 0.2)$~\kms\ is outside the \cite{2004ARA&A..42..685Z} ``good box'' of young stars, with $-15 \leqslant U \leqslant 0$~\kms, $-34 \leqslant V \leqslant -10$~\kms, and $-20 \leqslant W \leqslant +3$~\kms, suggesting an age of $\gtrsim 150$~Myr.

{\it HR Diagram Position.} Pre-main sequence M dwarfs with comparable masses as COCONUTS-2A are still in the process of contraction over their first $\sim 100$~Myr, and thus should have brighter absolute magnitudes than field-age objects. We find Gaia photometry of COCONUTS-2A lines up with the main sequence, as well as members of AB~Doradus ($\approx 149$~Myr) and older groups \citep[e.g.,][]{2021AJ....161..277K}, suggesting an age of $\gtrsim 150$~Myr.

Altogether, we adopt an age estimate of $150-800$~Myr for the COCONUTS-2 system.

\section{COCONUTS-2\lowercase{b}: The Cold Wide-Orbit Exoplanet}
\label{sec:planet}

\begin{figure*}[t]
\begin{center}
\includegraphics[height=6in]{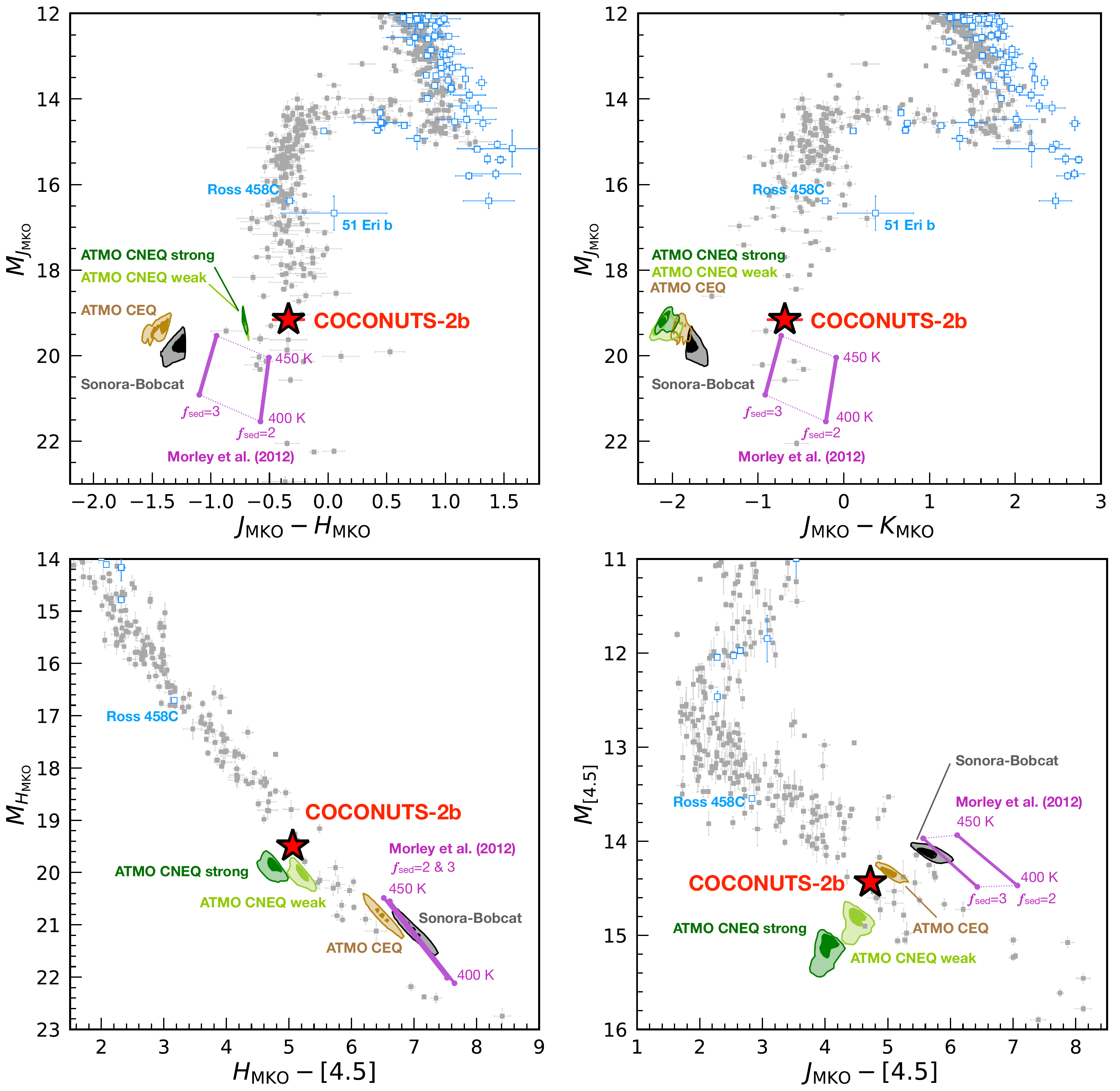}
\caption{Photometry of COCONUTS-2b (red star), overlaid with low-gravity L and T dwarfs in the field or young moving groups (blue open squares) and (high-gravity) field dwarfs that have absolute magnitudes and colors with S/N$>$5 and are not resolved binaries or subdwarfs (grey squares). Photometric uncertainties of COCONUTS-2b (see Table~\ref{tab:info}) are shown if they exceed the size of the symbol. We use contours to show $1\sigma$ (darker shade) and $2\sigma$ (lighter shade) confidence intervals of the predicted photometry from four sets of evolutionary models given the bolometric luminosity and age of COCONUTS-2b: Sonora-Bobcat (dark grey), ATMO CEQ (brown), ATMO CNEQ weak (light green), and ATMO CNEQ strong (dark green). We also plot the \cite{2012ApJ...756..172M} cloudy models (purple) with \logg$=4.0$, and \Teff$=400-450$~K, corresponding to COCONUTS-2b's properties (Section~\ref{sec:planet}). Tracks with condensate sedimentation efficiencies of both $f_{\rm sed} =2$ and $3$ are shown. Model-predicted photometry for the same $f_{\rm sed}$ are connected by solid lines and photometry by dotted lines for the same \Teff. COCONUTS-2b might have both clouds and non-equilibrium processes in its atmosphere, as expected given its cold effective temperature.}
\label{fig:phot}
\end{center}
\end{figure*}

COCONUTS-2b was initially found as a field dwarf by \cite{2011ApJS..197...19K}, who assigned a T9 spectral type based on low-S/N near-infrared spectra. \cite{2017ApJ...842..118L} suggested this object is likely a subdwarf by comparing its $[3.6] - [4.5]$ and $J_{\rm MKO}-[4.5]$ colors with trends in the multi-metallicity \cite{2015ApJ...804L..17T} model atmospheres (though changes in these models' adiabatic indices can also affect these colors and the models do not accurately reproduce these colors for most late-T and Y field dwarfs --- see Figure~8 of \citealp{2017ApJ...842..118L}). \cite{2019ApJS..240...19K} found this T9 dwarf has fainter absolute magnitudes in $H$, $W1$, and $[3.6]$ bands than other field dwarfs with similar $[3.6]-[4.5]$ colors but also noted such peculiarity is not a hallmark of subdwarfs. As shown here, COCONUTS-2b is in fact a bound companion to a young solar-metallicity star, eliminating the possibility of it being a subdwarf.   

To infer the physical properties of COCONUTS-2b, we first derive its bolometric luminosity using the \cite{2013Sci...341.1492D} super-magnitude method as updated by W. Best \etal\ (in preparation). Using polynomials determined from the Sonora-Bobcat model atmospheres (Marley \etal, submitted), we combine and convert the object's absolute magnitudes in $J_{\rm MKO}$, $H_{\rm MKO}$, $[3.6]$, and $[4.5]$ bands into \logLbol$=-6.384 \pm 0.028$~dex. To derive the companion's physical properties (e.g., effective temperature \Teff, surface gravity \logg, radius \R, and mass \M), we then use our measured bolometric luminosity (assumed to follow a normal distribution) and its host star's age (assumed to follow a uniform distribution within 150--800~Myr) to interpolate both the Sonora-Bobcat (Marley \etal, submitted) and the ATMO~2020 \citep{2020A&A...637A..38P} evolutionary models. Both these models adopt hot-start initial conditions and assume objects form with high initial entropy without any accretion. For ATMO~2020, we use all three available versions: chemical-equilibrium models (CEQ), and non-equilibrium models with two levels of eddy mixing ($K_{zz}$; CNEQ weak, and CNEQ strong). All these models are for solar metallicity, cloud-free atmospheres.  The Sonora-Bobcat and ATMO CEQ models assume rainout chemical equilibrium. We find the physical properties derived from all these evolutionary models are consistent within $1\sigma$, with the ATMO 2020 models giving \Teff$=434\pm9$~K, \logg$=4.11^{+0.11}_{-0.18}$~dex, \R$=1.11 \pm 0.03$~\Rjup, and \M$=6.3^{+1.5}_{-1.9}$~\Mjup. At such surface gravity, the ATMO CNEQ weak and strong models correspond to $\log{K_{zz}} \approx 5$ and $7$, respectively \citep[see Figure~1 of][]{2020A&A...637A..38P}. 

For COCONUTS-2b, hot-start initial conditions are vastly more likely than a cold-start via core accretion, as a finely-tuned dynamical kick would be needed to remove it from its starting point in the protoplanetary disk of COCONUTS-2A to its present location but not eject it. Regardless, for a point of comparison, we use the cold-start version of the Sonora-Bobcat models (Marley et al., submitted; also see Section~5.5 of \citealt{2019AJ....158...13N}) to characterize COCONUTS-2b and find the derived physical properties (e.g., \M$=6.5^{+1.7}_{-2.0}$~\Mjup) are all consistent with hot-start model predictions within $1\sigma$ (Table~\ref{tab:info}). We also derive COCONUTS-2b's properties from the cold-start models of \cite{2012ApJ...745..174S} using the same method as Dupuy et~al.\ (2021, submitted) did for 51~Eri~b. The \cite{2012ApJ...745..174S} models cover the most tabulated masses (1--15\,\Mjup) and ages (1\,Myr to 1\,Gyr) of any cold-start models, and they also provide a wide range of initial specific entropy ($\Sinit$) values.
The online \citet{2012ApJ...745..174S} model data\footnote{\url{https://www.astro.princeton.edu/~burrows/warmstart/index.html}} include a model spectrum at each grid point, which we numerically integrate to compute \Lbol\ at each mass and age. 
Using their cloud-free, solar-metallicity evolutionary models, we find $M=11.6^{+1.9}_{-2.8}$\,\Mjup\ 
from the coldest-start initial conditions ($\Sinit=8.0$\,\kbb), which is significantly higher than from hot-start Sonora-Bobcat and ATMO~2020 models, as well as the cold-start Sonora models.

Compared to (older) field T9 dwarfs \citep[e.g.,][]{2013Sci...341.1492D, 2021ApJS..253....7K}, COCONUTS-2b has $0.5-1.5$~mag fainter absolute magnitudes in $Y/J/H/K/[4.5]$ bands, a similar $M_{[3.6]}$, $\approx 0.3$~mag bluer $[3.6]-[4.5]$ color, and $\approx 0.5$~mag redder $J-[4.5]$ and $H-[4.5]$ colors. Also, COCONUTS-2b has $0.2-0.6$~dex fainter \Lbol, $100-250$~K cooler \Teff, and $\approx 1.0$~dex lower \logg\ than those of T9 field dwarfs as studied by \cite{2013Sci...341.1492D} using the \cite{2003A&A...402..701B} evolutionary models with an assumed age of 5~Gyr. These might suggest a surface-gravity dependence of ultracool dwarf properties at T/Y transition, similar to the well-known phenomenon for the L/T transition \citep[e.g.,][]{2006ApJ...651.1166M, 2016ApJ...833...96L}. All these peculiarities of COCONUTS-2b are similar to those of the T6.5 imaged exoplanet 51~Eri~b, but they are not seen for the T8 wide-orbit planetary-mass companion Ross~458C. High-quality spectroscopy and a more accurate spectral type of COCONUTS-2b would help investigate whether its properties are indeed abnormal compared to the field population, given that its current T9 spectral type is based on relatively low-S/N spectra \citep[][]{2011ApJS..197...19K}.

To investigate the atmospheric processes of COCONUTS-2b, we compare its infrared photometry with the predicted photometry by hot-start cloudless Sonora-Bobcat and ATMO~2020 models given this planet's bolometric luminosity and age (Figure~\ref{fig:phot}). We also compare with the \cite{2012ApJ...756..172M} equilibrium-chemistry models with  optically thin sulfide and salt clouds, which might provide a better match to late-T dwarfs' spectrophotometry than cloudless models \citep[e.g.,][]{2020arXiv201112294Z, 2021arXiv210505256Z}. We find the near-infrared photometry of COCONUTS-2b can be best explained by cloudy models with condensate sedimentation efficiencies of $f_{\rm sed} = 2$ or $3$. All cloudless models predict too blue near-infrared colors, although the ATMO non-equilibrium models perform significantly better than equilibrium models in predicting both $M_{J_{\rm MKO}}$ and $J_{\rm MKO} - H_{\rm MKO}$ of COCONUTS-2b. When mid-infrared magnitudes are considered, we find the photometry of COCONUTS-2b might be better explained by ATMO non-equilibrium models than cloudy models, with the latter models significantly off the field sequence of T/Y dwarfs \citep[also see discussion in][]{2017ApJ...842..118L}. COCONUTS-2b might have both clouds and non-equilibrium processes in its atmosphere, which is not surprising given its cold effective temperature. The coexistence of these atmospheric processes have also been suggested for Ross~458C based on spectroscopic analysis \citep[e.g.,][]{2020arXiv201112294Z}.

\begin{figure*}[t]
\begin{center}
\includegraphics[height=4.in]{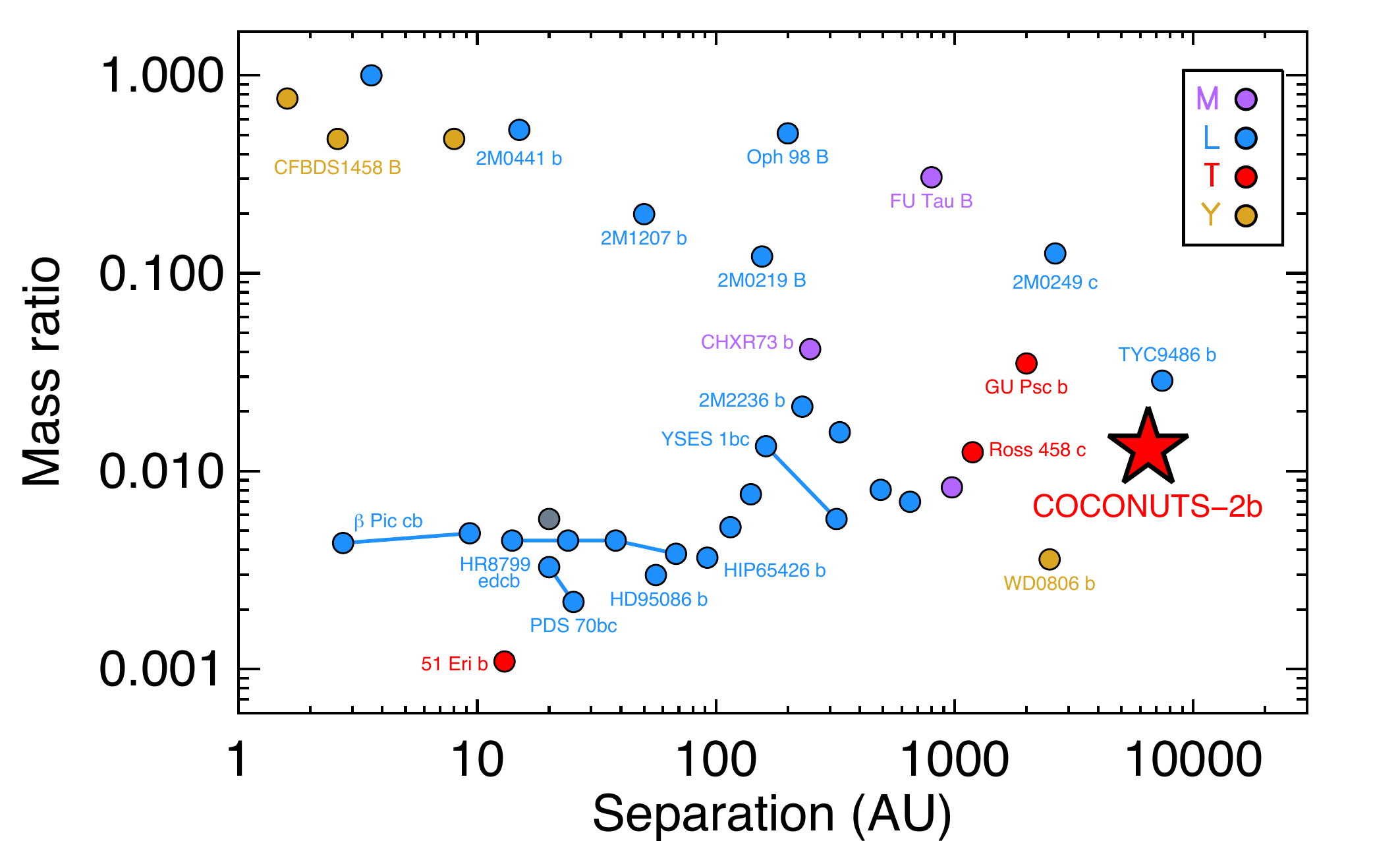}
\caption{Mass ratios and projected separations of planetary-mass ($<$15~\Mjup) companions, based on the compilation by \cite{2016PASP..128j2001B} with additions from The UltracoolSheet (and references therein), \cite{2020ApJ...898L..16B}, \cite{2020A&A...642L...2N}, \cite{2020ApJ...905L..14F}, \cite{2020AJ....159..263W}, and \cite{2021A&A...648A..73B}. Colors denote the spectral types of the companions (in some cases estimated from photometry).  Lines denote multi-planet systems.  COCONUTS-2b is the second widest \citep[after TYC~9486-927-1~b;][]{2016MNRAS.457.3191D} and the second coldest \citep[after WD~0806$-$661B;][]{2011ApJ...730L...9L} exoplanet imaged to date.}
\label{fig:massratio}
\end{center}
\end{figure*}

\section{Discussion}
\label{sec:discussion}

The COCONUTS-2 system has a mass ratio of $0.016^{+0.004}_{-0.005}$ based on the hot-start model-predicted mass of COCONUTS-2b (Figure~\ref{fig:massratio}). This ratio, combined with the companion's low mass of $6.3^{+1.5}_{-1.9}$~\Mjup, satisfy the IAU working definition of an exoplanet (which suggests an upper limit of $\approx0.04$ in mass ratio).\footnote{\url{https://www.iau.org/science/scientific_bodies/commissions/F2/info/documents/}} At 10.9~pc, COCONUTS-2b is the nearest imaged exoplanet to Earth to date. Given its projected separation of $6471$~au and $T_{\rm eff} \approx 434$~K, COCONUTS-2b is the second widest \citep[after TYC~9486-927-1~b at a projected separation of 6900~au;][]{2016MNRAS.457.3191D} and the second coldest \citep[after WD~0806$-$661B with \Teff$\approx 328$~K;][]{2011ApJ...730L...9L} imaged exoplanet discovered so far.  

\begin{figure*}[t]
\begin{center}
\includegraphics[height=5.5in]{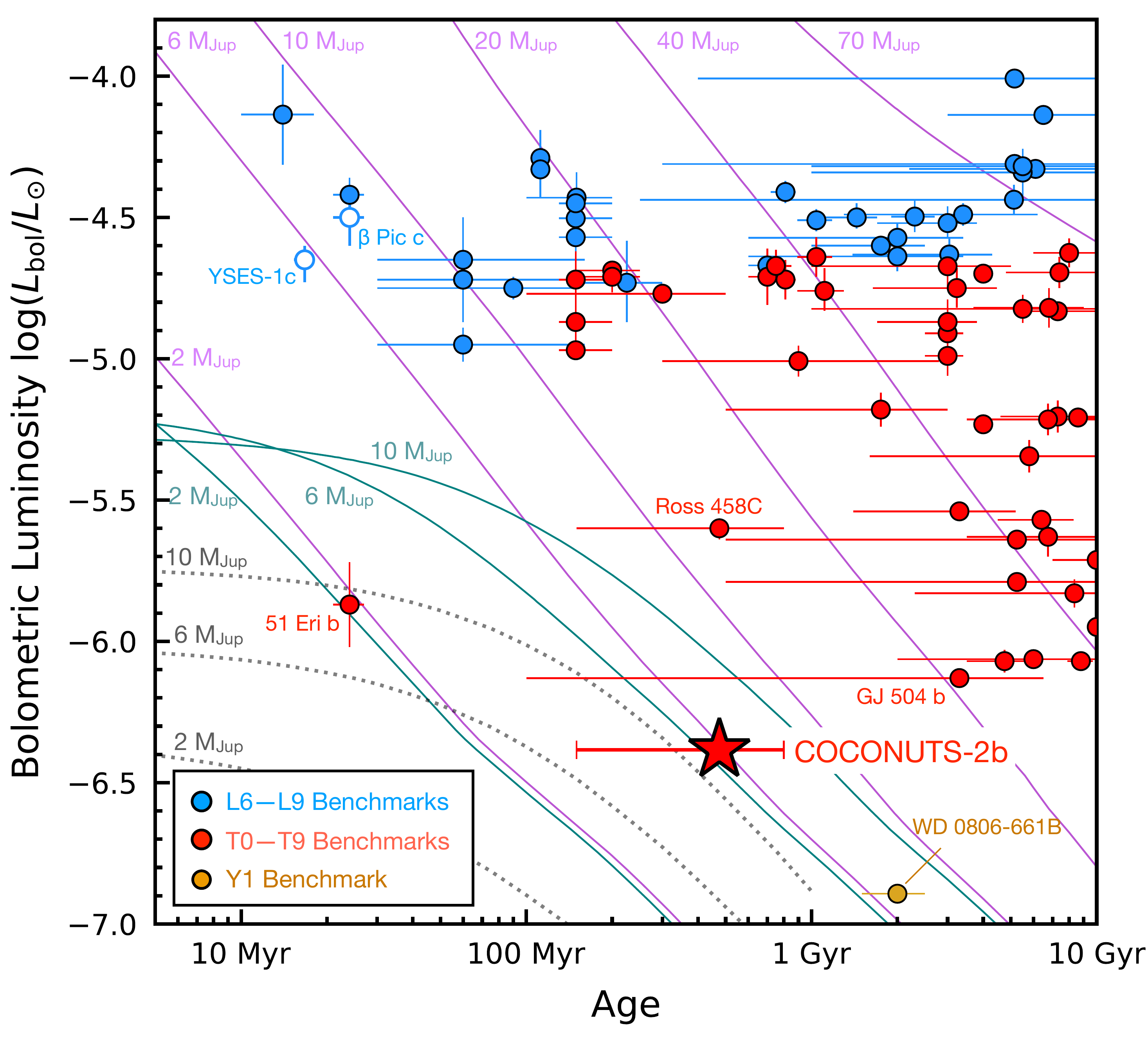}
\caption{Bolometric luminosity and age of COCONUTS-2b (red star) and those of L6$-$Y1 benchmarks from \cite{2020ApJ...891..171Z} and \cite{2021ApJ...911....7Z}. We also plot $\beta$~Pic~c \citep[][]{2020A&A...642L...2N} and YSES-1c \citep[][]{2020ApJ...898L..16B}, which have no spectral types but have bolometric luminosities consistent with late-L dwarfs.  We overlay the hot-start cloudless, solar-metallicity Sonora-Bobcat evolutionary models (purple solid; Marley et al., submitted) and the cold-start variation of these models (teal solid). We also show the coldest-start cloud-free, solar-metallicity \cite{2012ApJ...745..174S} evolutionary models (grey dotted), which are significantly fainter than the cold-start Sonora models. Similar to 51~Eri~b  \citep[][]{2015Sci...350...64M}, COCONUTS-2b has a sufficiently low luminosity to be consistent with the cold-start process that may form gas-giant (exo)planets, though its large separation means such formation would not have occurred in situ. }
\label{fig:lbol_age}
\end{center}
\end{figure*}

To convert COCONUTS-2b's projected separation into a semi-major axis, we use the scaling factor of $1.16^{+0.81}_{-0.32}$ from \cite{2011ApJ...733..122D}. We assume this factor follows a distribution composed of two half-Gaussians joined at 1.16, with a standard deviation of 0.81 and 0.32 toward higher and lower values (truncated at 0), respectively, leading to a semi-major axis of $7506^{+5205}_{-2060}$~au. Combining this value with the primary star's mass, we derive a very long orbital period of $1.1^{+1.3}_{-0.4}$~Myr for COCONUTS-2b. We also compute the binding energy of the COCONUTS-2Ab system as $(4.7^{+2.9}_{-1.9}) \times 10^{39}$~erg by using COCONUTS-2b's hot-start model-derived mass, which is the lowest among all host systems of imaged exoplanets. 

In addition, we note that COCONUTS-2A is not the brightest star in the night sky of COCONUTS-2b, especially given the wide orbital separation and the low luminosity of the primary star. Over time, the brightest star in the sky likely varies as the system drifts relative to bright stars in the solar neighborhood from the perspective of COCONUTS-2b. The star Canopus matches the flux of the host star COCONUTS-2A at 500~nm and exceeds its output for wavelengths bluer than $\sim 500$~nm. A human observer, if present, would likely perceive Canopus as the brightest source in the sky.

Figure~\ref{fig:lbol_age} compares the bolometric luminosity and age of COCONUTS-2b to previously known L6$-$Y1 planetary-mass and substellar benchmarks from \cite{2020ApJ...891..171Z} and \cite{2021ApJ...911....7Z}, as well as the hot-start and cold-start Sonora-Bobcat (Marley et al. submitted) and the coldest-start \cite{2012ApJ...745..174S} evolutionary models. Similar to 51~Eri~b \citep[][]{2015Sci...350...64M}, COCONUTS-2b has a sufficiently low luminosity to be consistent with the cold-start process that may form gas-giant (exo)planets. However, given its wide orbital separation, COCONUTS-2b probably formed in situ, like components in stellar binaries via the gravitational collapse of molecular cloud. In future work, high-quality spectroscopic follow-up and photometric monitoring of COCONUTS-2b can probe its atmosphere's composition (e.g., C/O ratio) and thermal structure, with the benefit of metallicity and abundance constraints provided from analysis of the host star.  This will allow us to investigate the formation mechanism of the COCONUTS-2Ab system and study the T/Y transition of self-luminous exoplanetary atmospheres at young ages for the first time.

\acknowledgements

We thank Jennifer van Saders for helpful discussions about stellar rotation and Eric Mamajek for helpful comments.
This work has benefited from The UltracoolSheet at \url{http://bit.ly/UltracoolSheet}, maintained by Will Best, Trent Dupuy, Michael Liu, Rob Siverd, and Zhoujian Zhang, and developed from compilations by \cite{2012ApJS..201...19D}, \cite{2013Sci...341.1492D}, \cite{2016ApJ...833...96L}, \cite{2018ApJS..234....1B}, and \cite{2020AJ....159..257B}.
This work has made use of data from the European Space Agency (ESA) mission
{\it Gaia} (\url{https://www.cosmos.esa.int/gaia}), processed by the {\it Gaia}
Data Processing and Analysis Consortium (DPAC,
\url{https://www.cosmos.esa.int/web/gaia/dpac/consortium}). Funding for the DPAC
has been provided by national institutions, in particular the institutions
participating in the {\it Gaia} Multilateral Agreement.
This paper includes data collected by the TESS mission, which are publicly available from the Mikulski Archive for Space Telescopes (MAST). Funding for the TESS mission is provided by NASA's Science Mission directorate. 
This research has made use of the Exoplanet Follow-up Observation Program website, which is operated by the California Institute of Technology, under contract with the National Aeronautics and Space Administration under the Exoplanet Exploration Program.
ZZ and MCL acknowledge support from the National Science Foundation through grant AST-151833.  This work is funded in part by the Gordon and Betty Moore Foundation through grant GBMF8550 to MCL.

\software{{\it emcee} \citep{2013PASP..125..306F}}

\end{CJK*}

\newpage
\bibliographystyle{aasjournal}
\bibliography{ms}

\clearpage
\tabletypesize{\scriptsize}

\begin{deluxetable}{lcllcll} 
\tablewidth{0pc} 
\setlength{\tabcolsep}{0.05in} 
\tablecaption{Properties of COCONUTS-2 \label{tab:info}} 
\tablehead{ \multicolumn{1}{l}{}  &  \multicolumn{1}{c}{}  &  \multicolumn{2}{c}{COCONUTS-2A}  &  \multicolumn{1}{c}{}  &  \multicolumn{2}{c}{COCONUTS-2b}  \\ 
\cline{3-4} \cline{6-7}  
\multicolumn{1}{l}{Properties}  &  \multicolumn{1}{c}{}  &  \multicolumn{1}{l}{Value}  &  \multicolumn{1}{l}{Ref.}  &  \multicolumn{1}{c}{}  &  \multicolumn{1}{l}{Value}  &  \multicolumn{1}{l}{Ref.}  }
\startdata
Spectral Type                          & &    M3                            & Torr06   & &    T9                         & Kirk11  \\
Age (Myr)                                & &   $150-800$                   & This Work   & &     --                          & $\dots$  \\
$[$Fe/H$]$ (dex)                     & &   $0.00 \pm 0.08$        & Hojj19   & &     --                          & $\dots$  \\
\hline
\multicolumn{7}{c}{Astrometry and Kinematics} \\
\hline
R.A., Decl. (epoch~J2000; hms, dms)       & &   07:49:12.68, $-$76:42:06.72               & Gaia16,20   & &    07:51:08.81, $-$76:34:49.43          & Kirk19    \\
$\mu_{\alpha}\cos{\delta}$, $\mu_{\delta}$ (mas/yr)   & &   $-102.15 \pm 0.02$, $-192.92 \pm 0.02$   & Gaia16,20   & &    $-104.8 \pm 2.8$, $-189.7 \pm 4.5$  & Kirk19    \\
Parallax (mas)                         & &   $91.826 \pm 0.019$   & Gaia16,20   & &   $97.9 \pm 6.7$    & Kirk19    \\
Tangential Velocity (km/s)        & &   $11.276 \pm 0.002$   & This Work   & &   $10.5 \pm 0.7$    & This Work    \\
Radial Velocity (km/s)              & &   $1.19 \pm 0.61$   & Schn19   & &   --    & $\dots$    \\
Position Angle (East of North; deg)  & &   --                & $\dots$   & &   $42.9$    & This Work    \\
Projected Separation  & &   --                & $\dots$   & &   $594''$ (6471~au)    & This Work    \\
\hline
\multicolumn{7}{c}{Spectrophotometric Properties} \\
\hline
$V$ (mag)        & &   $11.276 \pm 0.065$   & Kira12   & &   --    & $\dots$    \\
Gaia DR2 $G$ (mag)        & &   $10.146 \pm 0.002$   & Gaia16,18   & &   --    & $\dots$    \\
Gaia DR2 $BP$ (mag)        & &   $11.574 \pm 0.005$   & Gaia16,18   & &   --    & $\dots$    \\
Gaia DR2 $RP$ (mag)        & &   $9.009 \pm 0.003$   & Gaia16,18   & &   --    & $\dots$    \\
2MASS $J$ (mag)        & &   $7.406 \pm 0.021$   & Cutr03   & &   --    & $\dots$    \\
2MASS $H$ (mag)        & &   $6.860 \pm 0.030$   & Cutr03   & &   --    & $\dots$    \\
2MASS $K$ (mag)        & &   $6.579 \pm 0.018$   & Cutr03   & &   --    & $\dots$    \\
MKO $Y$ (mag)        & &   --   & $\dots$   & &   $20.020 \pm 0.100$    & Legg15    \\
MKO $J$ (mag)        & &   --   & $\dots$   & &   $19.342 \pm 0.048$    & Kirk11    \\
MKO $H$ (mag)        & &   --   & $\dots$   & &   $19.680 \pm 0.130$    & Legg15    \\
MKO $K$ (mag)        & &   --   & $\dots$   & &   $20.030 \pm 0.200$    & Legg15    \\
$W1$ (mag)              & &   $6.490 \pm 0.081$\tablenotemark{a}   & Cutr14   & &   $16.946 \pm 0.066$\tablenotemark{a}    & Cutr14    \\
$W2$ (mag)              & &   $6.244 \pm 0.027$   & Cutr14   & &   $14.530 \pm 0.036$\tablenotemark{a}    & Cutr14    \\
Spitzer/IRAC $[3.6]$ (mag)     & &   --   & $\dots$   & &   $14.416 \pm 0.036$    & Kirk11    \\
Spitzer/IRAC $[4.5]$ (mag)     & &   --   & $\dots$   & &   $14.620 \pm 0.020$    & Kirk11    \\
EW$_{\rm Li \lambda6708}$ (\r{A})     & &   $<0.05$   & Schn19   & &   --    & \dots    \\
EW$_{\rm H\alpha}$ (\r{A})     & &   $2.3-8.0$   & $\star$\tablenotemark{b}   & &   --    & \dots    \\
\hline
\multicolumn{7}{c}{Physical Properties} \\
\hline
$\log{(L_{H\alpha}/L_{\rm bol})}$ (dex)     & &   $-3.74 \pm 0.13$   & This Work   & &   --    & \dots    \\
$\log{(L_{X})}$ (dex)     & &   $28.92 \pm 0.02$   & Schn19   & &   --    & \dots    \\
$P_{\rm rot}$ (days)     & &   $2.83 \pm 0.28$   & This Work   & &   --    & \dots    \\
$\log{(L_{\rm bol}/L_{\odot})}$ (dex)    & &   $-1.732 \pm 0.011$  & This Work    & &    $-6.384 \pm 0.028$                                 & This Work    \\
\Teff\ (K)                                               & &   $3406 \pm 69$         & Gaid14    & &    hot-start: $434\pm9$ (ATMO), $429\pm9$ (Sonora-Bobcat)         & This Work    \\
 & &  & & &    cold-start: $431\pm9$ (Sonora-Bobcat)      & This Work    \\
\logg\ (dex)                                           & &   $4.83 \pm 0.03$       & This Work    & &    hot-start: $4.11^{+0.11}_{-0.18}$ (ATMO), $4.08^{+0.12}_{-0.18}$ (Sonora-Bobcat)        & This Work    \\
& & & & &    cold-start: $4.11^{+0.12}_{-0.19}$ (Sonora-Bobcat)         & This Work    \\
\R\ (A: R$_{\odot}$; b: R$_{\rm Jup}$)  & &   $0.388 \pm 0.011$   & This Work    & &    hot-start: $1.11 \pm 0.03$ (ATMO), $1.13^{+0.04}_{-0.03}$ (Sonora-Bobcat)         & This Work    \\
 & & &  & &    cold-start: $1.12^{+0.04}_{-0.03}$ (Sonora-Bobcat)         & This Work    \\
\M\ (A: M$_{\odot}$; b: M$_{\rm Jup}$) & &   $0.37 \pm 0.02$      & This Work    & &    hot-start: $6.3^{+1.5}_{-1.9}$ (ATMO), $6.1^{+1.5}_{-1.8}$ (Sonora-Bobcat)         & This Work    \\
& &  &  & &    cold-start: $6.5^{+1.7}_{-2.0}$ (Sonora-Bobcat), $11.6^{+1.9}_{-2.8}$ (Spie12)         & This Work    \\
\enddata 
\tablenotetext{a}{AllWISE photometry might be contaminated by the scattered light halo from a nearby bright source. } 
\tablenotetext{b}{EW$_{\rm H\alpha}$ was measured as $2.3-8.0$~\r{A} in literature \citep[e.g.,][]{2006AandA...460..695T, 2014MNRAS.443.2561G, 2019AJ....157..234S} with a median of $4.1$~\r{A}. } 
\tablerefs{Bail21: \cite{2021AJ....161..147B}, Cutr03: \cite{2003yCat.2246....0C}, Cutr14: \cite{2014yCat.2328....0C}, Gaia16: \cite{2016AandA...595A...1G}, Gaia18: \cite{2018AandA...616A...1G}, Gaia20: \cite{2020arXiv201201533G}, Gaid14: \cite{2014MNRAS.443.2561G}, Hojj19: \cite{2019AandA...629A..80H}, Kirk11: \cite{2011ApJS..197...19K}, Kira12: \cite{2012AcA....62...67K}, Kirk19: \cite{2019ApJS..240...19K}, Legg15: \cite{2015ApJ...799...37L}, Spie12: \cite{2012ApJ...745..174S}, Schn19: \cite{2019AJ....157..234S}, Torr06: \cite{2006AandA...460..695T}} 
\end{deluxetable}

\vfill
\eject
\end{document}